\documentclass[10pt,twocolumn,english,aps,showpacs,manuscript,prb,tightenlines]{revtex4}
\usepackage{times}
\usepackage[T1]{fontenc}
\usepackage[latin1]{inputenc}
\usepackage{amsmath}
\usepackage{babel}
\usepackage{graphics}
\usepackage{amssymb}

\makeatletter

\providecommand{\LyX}{L\kern-.1667em\lower.25em\hbox{Y}\kern-.125emX\@}

\makeatother
\begin{document}


\title{Effects of site dilution on the magnetic properties of geometrically
frustrated antiferromagnets}

\author{Angel J. Garcia--Adeva}

\email{garcia@landau.physics.wisc.edu}

\author{David L. Huber}

\affiliation{Department of Physics; University of Wisconsin--Madison; Madison,
WI 53706}

\begin{abstract}
The effect of site dilution by non magnetic impurities on the susceptibility
of geometrically frustrated antiferromagnets (kagome and pyrochlore
lattices) is discussed in the framework of the Generalized Constant
Coupling model, for both classical and quantum Heisenberg spins. For
the classical diluted pyrochlore lattice, excellent agreement is found
when compared with Monte Carlo data. Results for the quantum case
are also presented and discussed.
\end{abstract}
\maketitle
\emph{Introduction.-}~During recent years, geometrically frustrated
antiferromagnets (GFAF) have emerged as a new class of materials that
exhibit novel phases at low temperatures\cite{GINGRAS2000, SCHIFFER1996b, Ramirez1994, GARDNER1999, RAJU1999, MARTINHO2001, WILLS2000, Earle99, HAGEMANN2001}.
The elementary unit in these systems is based on the triangle, which
makes it impossible to satisfy all the antiferromagnetic interactions
at the same time, leading to frustration. Examples of GFAF are the
pyrochlore and kagome lattices. Experimentally, it is found that the
susceptibility in materials belonging to this class exhibits a high
temperature phase in which it follows the Curie--Weiss (CW) law. Below
the Curie--Weiss temperature, geometrical frustration inhibits the
formation of a long range ordered (LRO) state, and the system remains
paramagnetic, even though there are strong correlations between units.
This phase is universally present in these systems, and it is called
the cooperative paramagnetic phase. Finally, at a certain temperature,
\( T_{f} \), which depends on the particular material and, usually,
is well below the CW temperature, there appear non universal phases:
some of the systems remain paramagnetic down to the lowest temperature
reached experimentally\cite{GARDNER1999}, some of them exhibit non
collinear ordered states\cite{RAJU1999}, or even some of them form
a spin glass state\cite{MARTINHO2001, WILLS2000, Earle99}, even though
the amount of disorder in the structure is very small.

In the simplest theoretical description of GFAF, the spins on the
lattice are regarded as Heisenberg spins with only nearest neighbor
(NN) interactions. In this picture, it is predicted that the non trivial
degeneracy of the ground state inhibits the formation of a LRO state,
and the system remains paramagnetic down to zero temperature\cite{Reimers1991}.
However, due to the presence of frustration, the NN exchange does
not fix an energy scale on the problem, and any small perturbation
can break the non trivial degeneracy of the ground state and lead
to some kind of ordered state. Therefore, it is especially important
to incorporate these possible perturbations in any model that tries
to explain the low temperature phases of these systems. Examples of
perturbations present in real systems are next nearest neighbor (NNN)
interactions, small anisotropies, long range dipole--dipole interaction,
or dilution by non magnetic impurities, to cite some.
\begin{figure}[ht!]
{\centering \includegraphics{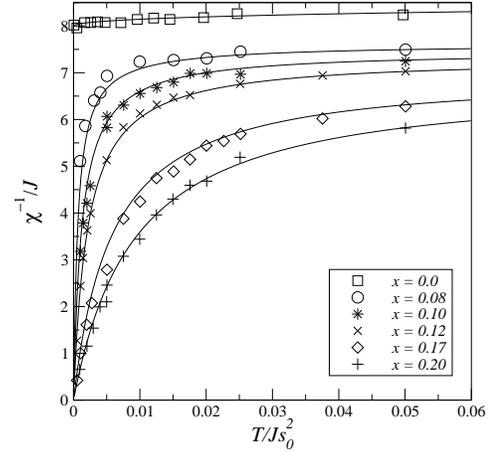} \par}

\caption{\label{fig.comparison.mc.pyro}Comparison between the inverse susceptibility
calculated in the GCC model for the diluted pyrochlore lattice for
different concentrations of non magnetic impurities \protect\( x\protect \)
and MC data from Ref.~\protect \onlinecite{Moessner99}.}
\end{figure}

Of course, before trying to understand the non universal behavior
observed below \( T_{f} \), it is very important to fully explore
models that \emph{quantitatively} describe the universal cooperative
paramagnetic phase. In a series of recent papers\cite{GARCIA-ADEVA2000, GARCIA-ADEVA2000a, GARCIA-ADEVA2000b},
the present authors have shown how very simple models based on small
clusters can provide a very accurate description of the universal
cooperative paramagnetic regime in these systems, when compared with
both Monte Carlo (MC) and experimental data. The purpose of the present
work is to study what are the effects of site dilution by non magnetic
impurities of the otherwise perfect pyrochlore and kagome lattices
on the magnetic properties in this same regime. This problem is relevant
for establishing comparisons with experimental data as, obviously,
any real material contains a certain amount of dilution. On more phenomenological
grounds, this problem was discussed in Refs.~\onlinecite{SCHIFFER1997}
and \onlinecite{Moessner99}.

\emph{The model.-}~The central idea of the models in Refs.~\onlinecite{GARCIA-ADEVA2000,GARCIA-ADEVA2000a,GARCIA-ADEVA2000b}
is based on the experimental and numerical observation that the correlations
in these systems are always short ranged. Therefore, we can start
by studying the properties of a small cluster and, later, add the
interactions with neighboring spins outside the cluster in an approximate
way. In the simplest approach, these interactions can be modeled as
a mean field. A more refined approximation consists on treating these
interactions as created by an effective field which is fixed by a
self consistency condition, the so called generalized constant coupling
(GCC) approach\cite{GARCIA-ADEVA2000a, GARCIA-ADEVA2000b}. This self
consistency condition is constructed in the following way: we consider
the magnetization per spin of a unit with \( p \) spins (a triangle
for the kagome lattice or a tetrahedron for the pyrochlore lattice,
for example), in the presence of the internal field created by the
neighboring \( p-1 \) spins outside the unit, and the magnetization
of an isolated spin in the presence of the internal field created
by the \( 2(p-1) \) neighboring spins, and we equate both quantities,
obtaining in this way an equation for the internal field \\
\begin{multline} 
m_{p}\left( \left[ H_{0}+(p-1)H'\right] /T\right) /p\\ 
=m^{CW}\left( \left[ H_{0}+2(p-1)H'\right] /T\right). 
\end{multline}\\
In this expression, \( H_{0} \) is the applied magnetic field and
\( H' \) is the internal field; \( m_{p} \) represents the magnetization
of a unit with \( p \) spins and \( m^{CW} \) the corresponding
magnetization for a single spin in the presence of the internal field
created by its \( 2(p-1) \) NN. Expressions for these quantities
can be found in references \onlinecite{GARCIA-ADEVA2000a} and \onlinecite{GARCIA-ADEVA2000b}
for the classical and quantum cases, respectively. Of course, this
equation can only be solved numerically in the general case but, in
the paramagnetic limit, \( \left[ H_{0}+2(p-1)H'\right] /T\ll 1 \),
we can expand it up to first order in the internal field in terms
of the susceptibility of an isolated unit and obtain an analytical
expression for the susceptibility per spin in the presence of the
internal field\begin{equation}
\chi _{p}^{gcc}(\widetilde{T})=\frac{1}{3\, J\, \widetilde{T}}\frac{1+\varepsilon _{p}(\widetilde{T})}{1-\varepsilon _{p}(\widetilde{T})},
\end{equation}
 where\begin{equation}
\label{epsilon.func}
\varepsilon _{p}(\widetilde{T})=\frac{2\widetilde{T}^{2}}{p}\frac{\partial }{\partial \widetilde{T}}\ln Z_{p}(\widetilde{T})-1,
\end{equation}
 with \( \widetilde{T} \) a dimensionless temperature defined by
\( T/\widetilde{J} \), where \( \widetilde{J}=Js_{0}^{2} \) in the
classical limit (\( J \) is the positive antiferromagnetic exchange
coupling and \( s_{0} \) the length of the spin) and \( \widetilde{J}=Js_{0}(s_{0}+1) \)
in the quantum case. \( Z_{p} \) is the partition function of an
isolated unit with \( p \) spins. Expressions for this quantity in
the classical limit can be found in Ref. \onlinecite{Moessner99}.
In the quantum limit, this quantity is given by\cite{GARCIA-ADEVA2000b}\begin{equation}
Z_{p}(\widetilde{T})=\sum _{S}g(S)(2S+1)e^{-\frac{S(S+1)}{2s_{0}(s_{0}+1)\widetilde{T}}},
\end{equation}
 where \( S \) represents the total spin of the unit and \( g(S) \)
the corresponding degeneracy. 

Let us now consider the effect of substituting some of the magnetic
ions in the lattice by non-magnetic ions. We will assume that the
distribution of non-magnetic impurities is completely random, so the
number of units with \( q=1,2,\ldots  \) magnetic ions for a lattice
formed by units with \( p \) ions, for a concentration of non-magnetic
impurities \( x \), is given by\cite{Moessner99}\begin{equation}
\mathcal{P}_{q}^{p}(x)=\tbinom {p}{q}\, (1-x)^{q}\, x^{p-q}.
\end{equation}

The self-consistency condition that determines the internal field
\( H' \) is given, in this case, in terms of the averaged magnetization
for different types of units \\
\begin{multline} 
\label{diluted.self.condition} 
(1-x)\, m^{CW}\left( \left[ H+2(p-1)(1-x)H'\right] /T\right) =\\
\frac{1}{p}\sum _{q=1}^{p}\mathcal{P}_{q}^{p}(x)\, m_{q}\left( \left[ H+(p-1)(1-x)H'\right] /T\right) , 
\end{multline}\\
where the \( (1-x) \) factor in front of the CW magnetization takes
into account the reduction of the total number of magnetic ions upon
dilution from \( N \) to \( N\, (1-x) \), whereas the prefactor
in the argument of both magnetizations stands for the reduction in
the number of NN.
\begin{figure}[th!]
{\centering \includegraphics{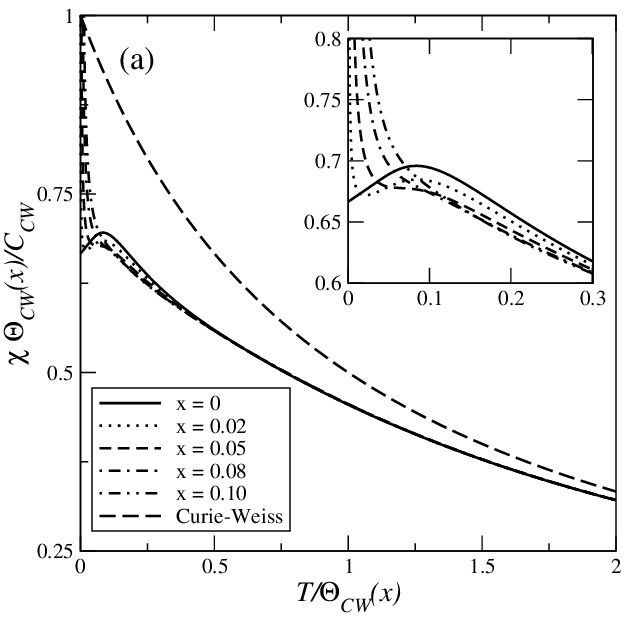} \par}

{\centering \includegraphics{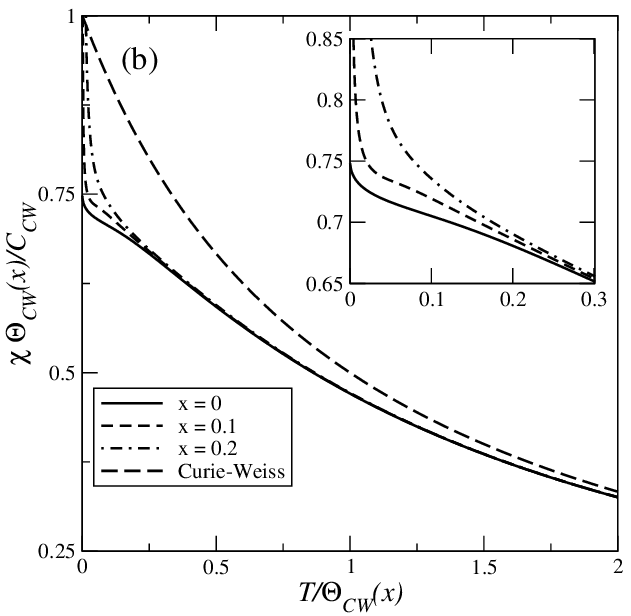} \par}

\caption{\label{fig.classical.suscep}Classical susceptibilities for the diluted
(a) kagome and (b) pyrochlore lattices. The inset shows the same quantity
at very low temperatures. The curve labeled as {}``Curie--Weiss{}''
is given by \protect\( 1/(1+T)\protect \) in this temperature scale.}
\end{figure}

Equation \eqref{diluted.self.condition} can be solved for \( H' \),
in the paramagnetic region, along the same lines mentioned above,
by introducing the function \begin{equation}
\overline{\varepsilon }_{p}(\widetilde{T},x)=\frac{2\widetilde{T}^{2}}{(1-x)p}\frac{\partial }{\partial \widetilde{T}}\ln \overline{Z}_{p}(\widetilde{T},x)-1,
\end{equation}
 which reduces to \eqref{epsilon.func} for \( x=0 \), where \( \overline{Z}_{p}(\widetilde{T},x) \)
is an averaged partition function defined by\begin{equation}
\overline{Z}_{p}(\widetilde{T},x)=\prod _{q=1}^{p}Z_{q}(\widetilde{T})^{\mathcal{P}_{q}^{p}(x)}.
\end{equation}

With these definitions, eq.~\eqref{diluted.self.condition} can be
solved in the paramagnetic regime to obtain the averaged susceptibility\begin{equation}
\overline{\chi }_{p}(\widetilde{T},x)=\frac{1-x}{3\, J\, \widetilde{T}}\frac{1+\overline{\varepsilon }_{p}(\widetilde{T})}{1-\overline{\varepsilon }_{p}(\widetilde{T})}.
\end{equation}

The corresponding internal energy per spin is given by\begin{equation}
\overline{u}_{p}(\widetilde{T},x)=\widetilde{J}\, \overline{\varepsilon }_{p}(\widetilde{T},x),
\end{equation}
 and the specific heat\begin{equation}
\overline{c}_{p}(\widetilde{T},x)=\frac{\partial }{\partial \widetilde{T}}\overline{\varepsilon }_{p}(\widetilde{T},x).
\end{equation}

\emph{Discussion.-}~The pyrochlore and kagome lattices have been extensively
studied by numerical methods in the classical limit\cite{Moessner99, Reimers1993b, Reimers1992c, Chalker92a}.
Therefore, there is a wealth of available data obtained from MC simulations
that allow us to estimate the accuracy of our model in the classical
limit. Actually, this comparison was recently carried out by the present
authors for the non diluted lattices considered in this work, and
it was found that both the susceptibility and specific heat are essentially
exact for the pyrochlore lattice down to 0 K. In the case of the kagome
lattice, however, the susceptibility was found to be again essentially
exact, but the MC specific heat deviates slightly as we go to 0 K,
due to the order by disorder effect which, obviously, is not included
in our model. We will not repeat such a comparison here, as it can
be found in Ref.~\onlinecite{GARCIA-ADEVA2000a}.
\begin{figure}[th!]
{\centering \includegraphics{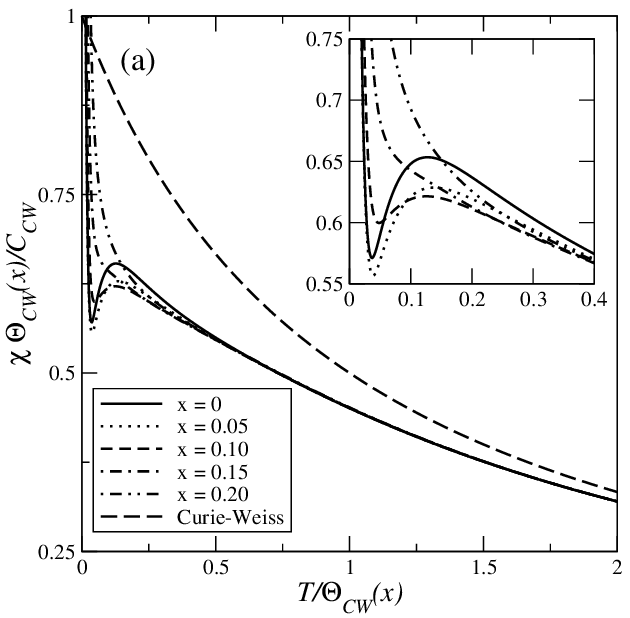} \par}

{\centering \includegraphics{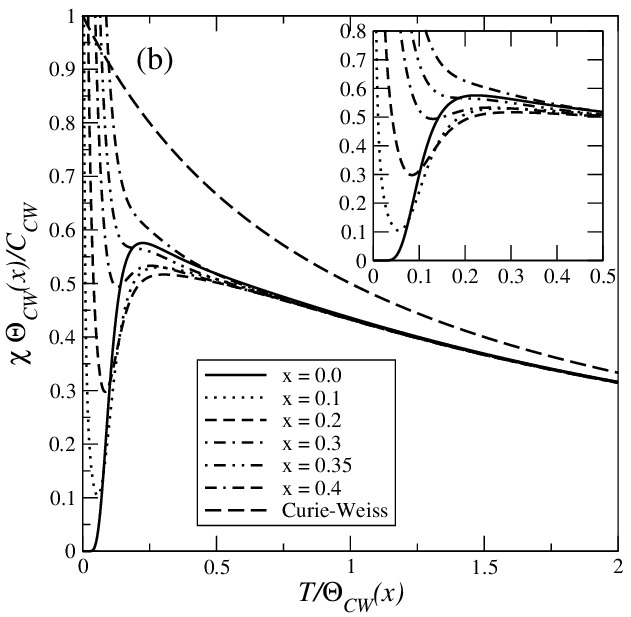} \par}

{\centering \includegraphics{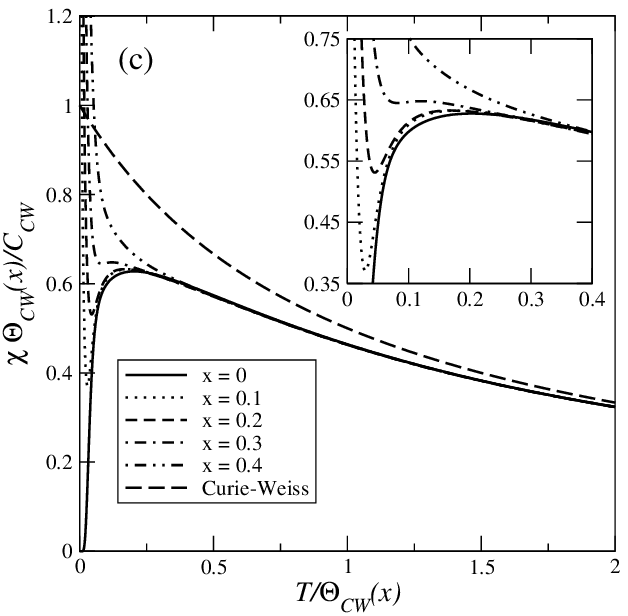} \par}

\caption{\label{fig.quantum.suscep}Quantum susceptibilities for the diluted
(a) kagome (\protect\( s_{0}=\frac{5}{2}\protect \)), (b) kagome
(\protect\( s_{0}=1\protect \)), and (c) pyrochlore (\protect\( s_{0}=\frac{3}{2}\protect \))
lattices. The inset shows the same quantity at very low temperatures.}
\end{figure}

There are also some results from MC calculations for the case with
site dilution but, unfortunately, only for the pyrochlore lattice\cite{Moessner99}.
They have been compared with the predictions of our model in the classical
limit in Fig.~\ref{fig.comparison.mc.pyro}. As can be seen, the susceptibility
predicted by our model is essentially exact down to 0 K. It is important
to notice that the curves in that figure do not include any adjustable
parameter or global scale factor. Of course, it would be extremely
interesting to test the predictions of the model for the kagome case
too.

In Figs.~\ref{fig.classical.suscep} and \ref{fig.quantum.suscep}
we have plotted the classical and quantum susceptibilities predicted
by our model for different concentrations of non magnetic impurities.
The quantum susceptibilities depicted there are for the kagome spin
5/2 and spin 1, and pyrochlore spin 3/2. The first one would correspond
to the iron jarosites compounds and the last one to ZnCr\( _{2} \)O\( _{4} \).
The second one would correspond to Mo\( ^{4+} \) ions on a kagome
lattice, and has been included to stress the differences between integer
and half integer values of \( s_{0} \) in our model. In these figures,
we have rescaled the temperature by the CW temperature\cite{SPALEK1986}
\( \Theta _{CW}(x)=\frac{2(p-1)\widetilde{J}}{3}(1-x) \), and the
susceptibility by the CW constant \( C_{CW}(x)=\frac{\widetilde{J}(1-x)}{3J} \).
The advantage of this rescaling is that all the curves for different
concentrations fall on the same curve at high temperatures, which
makes it easier to identify the nontrivial effects of dilution at
low temperatures (that is, those which do not come from a reduction
of \( (1-x) \) factor in both the CW temperature and CW constant).

From those figures, we can see that the main effect of dilution in
the quantum kagome and pyrochlore systems, and also in the classical
kagome lattice is that the intensity of the maximum in the susceptibility
decreases as the concentration of impurities increases and, eventually,
this maximum disappears for a certain amount of non magnetic impurities.
In the classical pyrochlore lattice, there is no maximum in the susceptibility,
but only a shoulder, due to the formation of short range correlations
in the tetrahedral unit. Again, this shoulder disappears for a certain
amount of non magnetic impurities. Another effect of dilution is related
to the upturn of the susceptibility at low temperatures. Both the
classical non diluted kagome and pyrochlore lattices susceptibilities
go to a finite value at 0 K. In the quantum case, the pyrochlore susceptibility
in the pure system falls to zero at 0 K. The kagome susceptibility
falls to zero for integer values of the individual spins or diverges
for half integer values of \( s_{0} \), even in the absence of dilution.
This divergence is related to considering a cluster with an odd number
of half integer spins, which has a ground state with a \( \frac{1}{2} \)
total spin momentum. This problem of the model and its implications
has been discussed in Ref.~\onlinecite{GARCIA-ADEVA2000b}. When we
add a certain amount of dilution, there is a divergence of the susceptibility
as we approach \( T=0 \) for all the cases considered in this work.
This upturn in both the classical kagome and pyrochlore lattices,
and also in the quantum pyrochlore and quantum kagome lattices with
integer values of \( s_{0} \), is due to the appearance of loose
spins in the system. However, in the quantum kagome with half integer
values of \( s_{0} \), this upturn is present even in the pure system,
as commented above. In the quantum pyrochlore lattice with half integer
\( s_{0} \), both the loose spins and triangular units contribute
to this upturn of the susceptibility.

Even though we will not present a detailed comparison with experimental
measurements here, it is interesting to note that some of the features
exhibited by the calculated quantum susceptibilities are qualitatively
similar to those experimentally found in pyrochlore systems\cite{MARTINHO2001, RAJU1999, GARDNER1999},
iron jarosites\cite{WILLS2000, Earle99} and also in SrCr\( _{9p} \)Ga\( _{12-9p} \)O\( _{19} \)\cite{MENDELS00}
(where these last two systems are considered as experimental realizations
of the kagome lattice), namely, a maximum in the susceptibility for
both the pyrochlore and kagome lattices. The disappearance of this
maximum with increasing dilution and the upturn of the susceptibility
below the maximum for pure samples has been also observed in the iron
jarosites compounds. However, a lot of caution has to be exercised
when comparing the present model with experimental data for two main
reasons: first, the region where the maximum of the susceptibility
is observed, it is precisely the region where spin freezing starts
to take place and, obviously, this effect is not accounted for in
our model. Second, it is not obvious that the distribution of non
magnetic impurities in real systems is purely random and this fact
could lead to important quantitative differences when trying to use
this model to interpret experimental data.

In any case, we think that the present model could provide further
insight in the properties of the geometrically frustrated antiferromagnets
in the cooperative paramagnetic region.

\begin{acknowledgments}
A.J.G.-A. wishes to acknowledge financial support from the Spanish
MEC under the Subprograma General de F.P.I. en el Extranjero.
\end{acknowledgments}

\end{document}